\documentclass[reprint,showpacs,preprintnumbers,prl,10pt,aps,groupedaddress]{revtex4-1}
\usepackage[pdftex]{graphicx}
\usepackage{dcolumn}
\usepackage{bm}
\usepackage{epsfig}
\usepackage{latexsym}
\usepackage{amsmath}
\usepackage{color}
\usepackage{array}

\newcommand{\<}{\langle}

\renewcommand{\>}{\rangle}
\renewcommand{\(}{\left(}
\renewcommand{\)}{\right)}
\renewcommand{\[}{\left[}
\renewcommand{\]}{\right]}
\renewcommand{\v}[1]{\mathbf{#1}} % \v -> vector (bf)
\newcommand{\dslash}{d \hspace{-0.8ex}\rule[1.2ex]{0.8ex}{.1ex}}

\newcommand{\sgn}{\text{sgn}}

\begin{document}
\title{Quantum spin liquids and the metal-insulator transition in doped semiconductors}
\author{Andrew C. Potter$^1$, Maissam Barkeshli$^2$, John McGreevy$^1$, and T. Senthil$^1$}
\affiliation{$^1$Department of Physics, Massachusetts Institute of Technology,
Cambridge, MA 02139, USA }
\affiliation{$^2$Department of Physics, Stanford University, Stanford, CA 94305, USA}

\date{\today}
\begin{abstract}
We describe a new possible route to the metal-insulator transition in doped semiconductors such as Si:P or Si:B. We explore the possibility that the loss of metallic transport occurs through Mott localization of electrons into a quantum spin liquid state with diffusive charge neutral ``spinon" excitations. Such a quantum spin liquid state can appear as an intermediate phase between the metal and the Anderson-Mott insulator. An immediate testable consequence is the presence of metallic thermal conductivity at low temperature in the electrical insulator near the metal-insulator transition. Further we show that  though the transition is second order the zero temperature residual electrical conductivity  will jump as the transition is approached from the metallic side. However the electrical conductivity will have a non-monotonic temperature dependence that may complicate the extrapolation to zero temperature.  Signatures in other experiments and some comparisons with existing data are made. 

\end{abstract}

\maketitle 

Phenomena near the metal-insulator transition (MIT) in doped semiconductors such as Si:P or Si:B have been studied extensively for more than three decades\cite{BhattReviews,BelitzRMP,LeeRMP,LohneysenReview}. Nevertheless, several aspects of the physics, for instance the detailed critical behavior\cite{QCScaling,QCScalingControversy,LohneysenReview},  remain mysterious.  In this paper we explore and develop the consequences of a new possible route to the MIT where a quantum spin liquid insulator appears as an intermediate phase between the metal and the Anderson-Mott insulator. In recent years such a quantum spin liquid Mott insulator has been observed to intervene between the Fermi liquid metal and conventional magnetically ordered Mott insulators in a few different clean materials\cite{OrganicTriangularLattice,DMITThermal,HyperKagome}. Here we study the strongly disordered situation appropriate to doped semiconductors and describe a variety of experimental consequences. 

When P is doped into Si, the extra electron of P forms a Hydrogen-like state with an effective Bohr radius of about $a\approx 20$~\AA \cite{BhattReviews}.  A simple picture of the doped semiconductor is as a 
collection of randomly placed ``Hydrogen" atoms.  The system may then be described as a half-filled Hubbard model on a random lattice supplemented by the inclusion of the long range Coulomb interaction $V_{ij}$ between the electrons: 
\begin{equation} \label{eq:Model}
H = -\sum_{ij; \alpha}t_{ij} \left(c^\dagger_{i\alpha} c_{j\alpha} + h.c \right) + U\sum_{i} n_{i\uparrow} n_{i\downarrow} + \sum_{i \neq j} V_{ij}n_in_j
\end{equation}
At low concentrations, the $t_{ij} \approx t_0 e^{-r_{ij}/a}$ are small, the on-site $U$ dominates and a Mott insulator of local moments results.  The local-moments are coupled antiferromagnetically, and due to their random placement, preferentially form singlets with their closest available neighbor.  The resulting random-singlet phase has an extremely broad distribution of singlet binding energies, giving rise to a diverging density of states for low-energy spin-excitations, contributing a divergent coefficient of heat capacity, $\gamma=C/T$ and spin-susceptibility $\chi$\cite{RandomSinglet}.

As the concentration of dopants, $n$, is increased, eventually the typical $t_{ij}$ dominates over the $U$ and a diffusive metal is obtained.  A continuous phase transition between metal and insulator occurs at some critical intermediate concentration, $n_c$, where $t_{ij}\approx U$.  Because of the random placement of dopants, a fraction of the local moments are very weakly coupled to the conducting electrons and survive unscreened into the metallic phase.   The diffusive metal appears to be well described by a ``two-fluid" model where the conducting electrons exist essentially decoupled from a random fraction of weakly-coupled local-moments\cite{BhattReviews,LMs}. As in the insulating phase, these local moments continue to dominate the low-temperature thermodynamic and magnetic properties of the metallic phase, but do not appear to strongly modify its transport properties.

\begin{figure}[ttt]
\begin{center}
\includegraphics[width=2.8in]{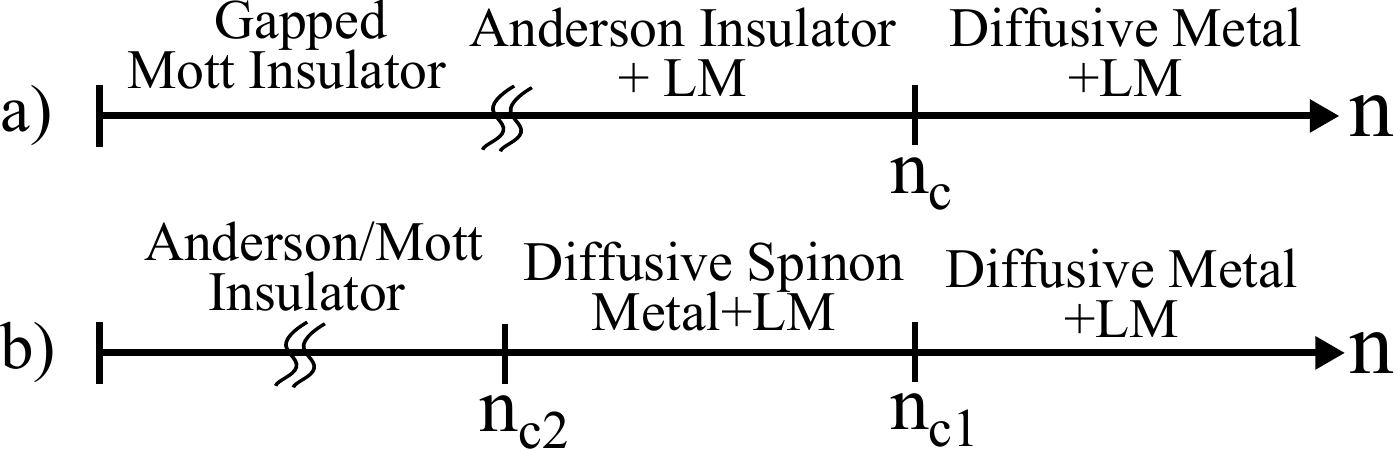}
\end{center}
\vspace{-.2in}
\caption{Two scenarios for the MIT in doped semiconductors.  (a) Conventional picture electron localization transition to Anderson-Mott insulator, which crosses over continuously to a pure Mott insulator (indicated by wiggly lines).  (b) In the newly proposed scenario, the transition is to a spin-liquid with gapless fermionic spinon excitations, here the electrical MIT and spin/thermal MIT  occur separately at  $n_{c1}$ and $n_{c2}$ respectively. ``+LM" indicates the fraction of randomly decoupled local spin-moments that inevitably accompany all phases.}
\label{fig:Scenarios}
\vspace{-.1in}
\end{figure}

It is natural to ask: What is the fate of the conducting fluid across the metal--insulator transition?  The conventional answer, implicitly adopted by most existing work\cite{BelitzRMP,LeeRMP}, is that all electron degrees of freedom are localized by disorder\cite{AndersonLocalization}, which is perturbatively enhanced by interactions. In this scenario, shown in Fig. \ref{fig:Scenarios}a, decreasing $n<n_c$ gives a localized Anderson-Mott insulator with non-zero average density of states. As $n$ is further decreased, the system crosses over continuously towards a correlation driven Mott-insulator of local moments. 

In this paper, we point out a new and conceptually distinct scenario for the metal--insulator transition in doped semiconductors.  In this scenario, the charged--conducting fluid is localized into a gapless quantum spin-liquid, but the electron thermal transport remains diffusive into the weakly insulating regime.  There is growing theoretical and experimental evidence that such gapless spin-liquids occur in clean Mott insulators, where strong charge fluctuations and frustration prevent magnetic ordering\cite{OrganicTriangularLattice,DMITThermal,HyperKagome}.  This experience makes it natural to ask whether or not one should expect a spin-liquid phase to form in (uncompensated) doped semiconductors near the MIT where charge fluctuations are strong, the system is at half-filling, and magnetic order is prevented by the random lattice structure, the competition between antiferromagnetic direct-exchange and random-sign RKKY exchange, and by quantum fluctuations.

\textit{Nature of the Possible Spin--Liquid Phase(s)} -- The proposed spin-liquid phase is most conveniently described by formally dividing the electron into a bosonic U(1) rotor $e^{i\theta}$ that carries the electron--charge, and a fermionic spinon $f_\alpha$ that carries the electron--spin: $c_i = e^{i\theta_i}f_{i\alpha} \label{eq:SlaveRotor}$\cite{FlorensGeorges,Senthil08}.  This description allows extraneous unphysical states that must be removed by constraining $n_{b,i}-\sum_\alpha f_{i\alpha}^\dagger f_{i\alpha}=1$ on each site, i.  Here $n_{bi}$ is the number operator conjugate to $\theta_i$.  The above decomposition has a U(1) gauge redundancy associated with $\theta_i \rightarrow \theta_i+\Lambda_i$ and $f_i\rightarrow e^{-i\Lambda_i}f_i$, which manifests itself in the low-energy effective theory as an emergent U(1) gauge field, $\v{a}(\v{r},t)$ \cite{LeeLee}.

Decoupling the hopping term $-t_{ij}c_{i,s}^\dagger c_{j,s} = -t_{ij}e^{i(\theta_i-\theta_j)}f_{i,s}^\dagger f_{j,s}$ in a mean field approximation and including gauge fluctuations gives the following effective action:  $S_{\text{eff}} = \int d\tau \(L_b+L_f\)$
\begin{eqnarray} \label{eq:MFAction}
&L_b = \sum_i \frac{1}{2}\(\partial_\tau\theta_i+a^0_i\)\(U\delta_{ij}+V_{ij}\)^{-1}\(\partial_\tau\theta_j+a^0_j\)
\nonumber\\
&\hspace{.25in}-t_{ij}\sum_{ij}\chi^f_{ij}e^{i(\theta_i-\theta_j+a_{ij})}
\nonumber \\
& L_f = \sum_{ij} f^\dagger_i\[\(\partial_\tau-a^0_i-\mu\)\delta_{ij} -t_{ij}\chi^b_{ij}e^{-ia_{ij}}\]f_{j,s} 
\end{eqnarray}
where $\chi^f_{ij} = \<f^\dagger_{i,s}f_{j,s}\>$ and $\chi^b_{ij} = \<e^{i(\theta_i-\theta_j)}\>$ are determined self-consistently. Note that, due to the random placement of sites, in general $\sum_{s}\<f^\dagger_{i,s}f_{i,s}\>$ will be spatially varying.  Consequently, even at the mean-field level, the bosons will experience a random chemical potential; this changes the universality class of the Bose-Mott transition compared to the clean case (where $\<n_b\> =1$ for every site on both sides of the Mott transition).

The metallic Fermi-liquid state corresponds to a superfluid--ordered phase of the bosonic rotors with $\<e^{i(\theta(\v{x})+\int^{\v{x}} \v{a}(\v{y})\cdot d\v{y})}\>\neq 0$, coexisting with a diffusive Fermi-liquid of spinons. In this phase, the emergent gauge field is gapped by the condensate of charged rotors through the Anderson--Higgs mechanism, and the rotors and spinons are ``glued" together into ordinary electrons.  

Eq. \ref{eq:MFAction} also naturally describes a deconfined state in which the rotors form an insulator, while the spinons remain diffusive.  This results in an exotic charge--insulator with finite-density of states for gapless spin-$1/2$ excitations.  We suggest that this spin-liquid phase may occur near the MIT for doped semiconductors.  In this scenario, shown in Fig. \ref{fig:Scenarios}b, the magnetic properties of the system change only gradually across the MIT at $n_{c1}$, and are qualitatively identical in both the metal and insulator.  In particular, we expect that one would still find a decoupled fraction of local-moments.  As these local moments dominate the low-temperature thermodynamics and magnetic properties, the clearest signature of the spin-liquid is metallic thermal conductivity, $\kappa\sim T$ at low $T$\cite{DMITThermal}.  While there has been extensive experimental analysis of the conductivity of doped semiconductors near the MIT, very little is known about thermal--transport.

In the slave-rotor language, the formation of local moments comes from rare strong fluctuations in disorder that locally bind the rotor and fermion back into a correlation--localized electron.  We assume that the principal effect of correlated disorder among the rotor, spinon, and gauge-field sectors is to produce such local moments, and that the physics of the remaining non-local moment bulk can be well described by treating disorder separately in each sector.

In the spin-liquid phase, the emergent gauge field is deconfined, and in clean systems its fluctuations lead to singular self-energies for the spinons resulting in non-Fermi-liquid behavior (2D)\cite{LeeGaugeTheory,Polchinski:1993ii,Nayak:1993uh,HLR,altshuler-1994,SSLeeLargeN,MetlitskiIsingNematic,MrossNEpsilon} or marginal Fermi-liquid behavior (3D)\cite{Podolsky,Holstein,Reizer}. For the strongly disordered doped semiconductors, the inelastic scattering rate for the spinons from gauge fluctuations scales as $\tau_{g}^{-1}\sim T$ and is dominated by the elastic impurity scattering for low $T$\cite{Supplement}. Consequently, the low-energy properties of the disordered spinon Fermi-liquid will be largely unaltered by the emergent gauge field. Furthermore, the gauge field propagation is strongly damped by the diffusive spinons, leading to an $\omega\sim q^2$ scaling of gauge-excitations.  This scaling implies that the gauge-field contribution to thermodynamic quantities is sub-dominant compared to the spinon contribution. For example the gauge-field specific heat scales as $C_a\sim T^{3/2}\ll C_{\text{spinon}}\sim T$.  

In 2D, a deconfined phase for the gauge-field requires the presence of extended, gapless fermionic excitations to suppress instanton configurations \cite{Polyakov,Hermele}.  In 3D, however, a compact U(1) gauge--field may remain deconfined even without extended, gapless matter \cite{Polyakov}.  Therefore, in addition to the gapless, thermally--conducting spinon Fermi-surface state described above, it is also possible to form an insulating state where the charge degrees of freedom are Mott localized and the spinons are Anderson-localized by disorder.  Such a spinon Anderson insulator is distinguished (in principle) from the conventional Anderson-Mott insulator by the presence of a gapless emergent U(1) gauge-field (though experimentally detecting the emergent gauge-field would be challenging).

\begin{figure}[ttt]
\begin{center}
\includegraphics[width=2.5in]{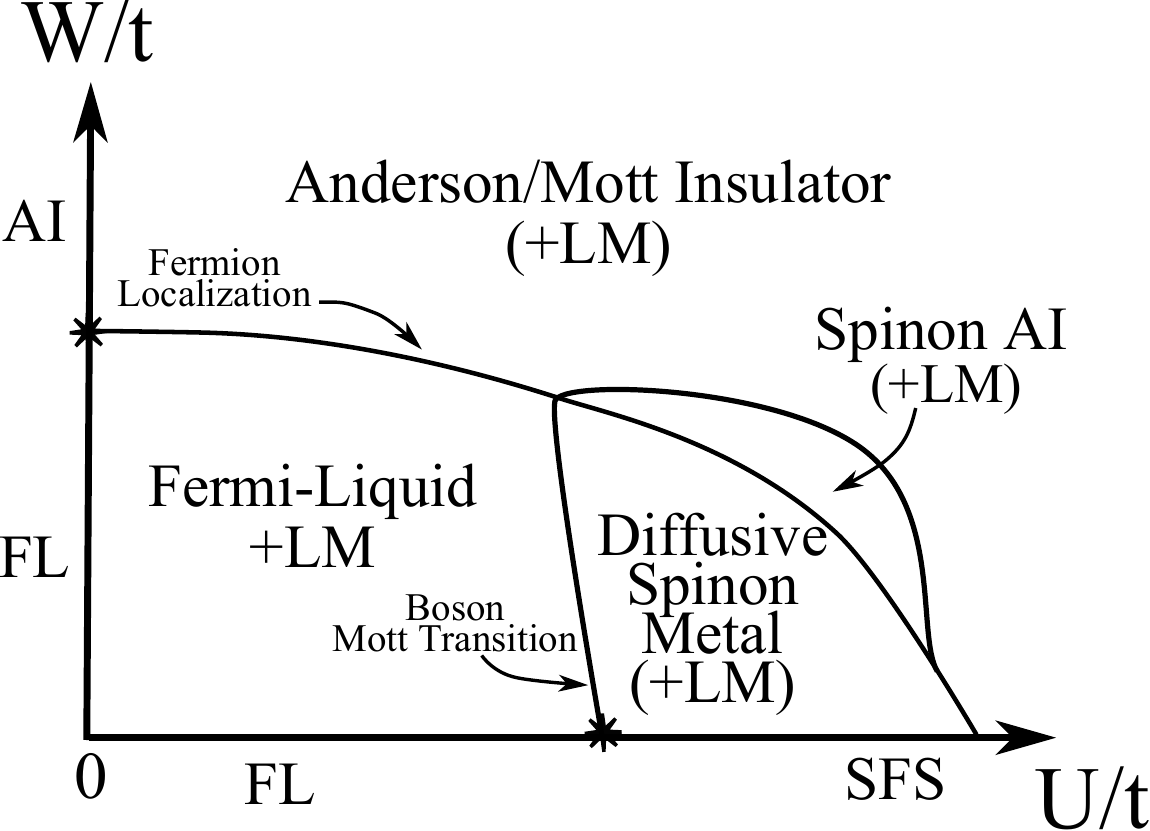}
\end{center}
\vspace{-.2in}
\caption{Schematic generalized phase--diagram as a function of disorder strength $W$ and interaction strengths $U$ measured with respect to the typical hopping $t$.}
%\caption{Schematic phase diagram in terms of disorder strength $W$, and ration of interaction strength $U$ to hopping $t$.  FL -- Fermi-Liquid, AI -- Localized Anderson Insulator, LM -- Local Moments (some dilute tail of local moments are present in all phases with non-zero $W$ and $U$), SFS -- Sharp Spinon Fermi Surface (non-Fermi-liquid state due to gauge fluctuations), SFL -- Spinon Fermi Liquid, SAI -- Spinon Anderson Insulator.  The AI at very strong disorder crosses-over continuously to a Mott-insulator of local moments, which order magnetically (forming an AFM at weak disorder, and likely transitioning to a spin-glass or random-singlet phase for stronger disorder).}
\label{fig:PhaseDiagram}
\vspace{-.1in}
\end{figure}

\textit{Generalized Phase Diagram} -- The MIT achieved by changing $n$, though experimentally relevant, is conceptually complicated since disorder and interactions are simultaneously affected.  It is conceptually simpler to consider a generalized phase-diagram where disorder strength $W$ and interaction strength $U$ can be separately adjusted, as in Fig. \ref{fig:PhaseDiagram}.  Here we restrict our attention to 3D, half-filling, and non-nested Fermi-surfaces (which are not inherently unstable to magnetic ordering). Furthermore, we remain agnostic about the particular realization of disorder, with the expectation that such details will not alter the qualitative discussion that follows.

We begin by considering various limiting cases.  The $(U = 0, W\neq 0)$ limit is  completely understood\cite{AndersonLocalization}: here a diffusive Fermi-liquid occurs up until a critical disorder strength beyond which all states near the Fermi-energy become localized leading to an Anderson insulator (AI).  Each of these phases is known to be stable to infinitesimal interactions, and therefore extends at least to small $U$.  The limit of $(U\neq 0, W\rightarrow \infty)$ is also straightforward.  Here the Anderson localized insulator at weak interactions crosses over continously to the Mott insulator of local-moments at strong-interactions.  At $T=0$, the local moments are magnetically ordered in either a random--singlet or spin-glass phase. 

Finally, the line $(U\neq 0, W = 0)$ is also reasonably well understood\cite{Senthil08}, albeit with slightly less confidence.  The clean Fermi-liquid survives up until some critical interaction strength, beyond which it becomes a weak Mott-insulator with a spinon Fermi surface (SFS).  For large $U$, the emergent gauge field undergoes a confinement transition and anti-ferromagnetic order develops.  Here again, each of the clean interacting phases is stable to infinitesimally small amounts of disorder and extends to finite $W$. The only distinction being that, for any $(U\neq 0, W\neq 0)$, disorder creates a non-zero density of decoupled local moments (indicated in Fig. \ref{fig:PhaseDiagram} by ``+LM").

These considerations greatly constrain the structure of the generalized phase--diagram.  Each of the phases at the boundary are known to extend to finite values of $W$ and $U$. Given the understanding of the boundaries of the phase diagram, the main issue here is not whether a strongly disordered fermionic spin-liquid could exist, but rather which particular path through the generic $W$ and $U$ phase--diagram is appropriate to tuning $n$ in doped semiconductors.  Fig.~\ref{fig:PhaseDiagram} depicts an extension of the well-understood outer boundary of the phase diagram to the interior.  While one can conceive of many intermediate insulating phases at intermediate $U$ and $W$, in the slave-rotor language, the only other natural candidate is the deconfined spinon Anderson insulator described above.

\textit{Thermal Conductivity} -- In the spin-liquid scenario, the electrical MIT and thermal MIT occur separately: whereas electrical conductivity vanishes in the insulating phase, thermal conductivity remains metallic, scaling as $\kappa_{\text{sp}}\sim T$ at low temperatures.  Since the ever-present concentration of local-moments dominates the low-temperature thermodynamical properties of the system (but contributes only weakly to transport), linear-T metallic conductivity is the clearest experimental signature of the spin-liquid. 

While $\kappa_{\text{sp}}\sim T$ at the lowest temperatures there will be Altshuler-Aronov--type corrections to $\kappa$ from interactions and disorder: $\kappa_{\text{AA}}\sim T^{3/2}$ \cite{Aleiner,Smith,Livanov}. Also, one expects a large contribution, $\kappa_{\text{ph}}$ from phonons: $\kappa_{\text{ph}}\sim T^3$ \cite{RSKappa}.  Therefore, to observe the metallic spinon-contribution, it may be necessary to work at very low temperature, and carefully subtract sub-dominant contributions. 

%One also may worry that the large density of low-energy excitations of the spectator local-moments, which dominate thermodynamic measurements, may also be important for $\kappa$. However, the local moments appear to form a random--singlet phase near the MIT indicating that these abundant low-energy excitations are localized and hence contribute only weakly to thermal transport.

%Finally, the ever-present fraction of localized spin-moments are strongly disordered, and likely have a large number of low--energy excitations that could potentially contribute to transport.  For the case of a spin--glass, phonon scattering off of these low energy states is empirically known to alter the phonon $\kappa_{\text{ph}}\sim T^2$.  However, for doped semiconductors like Si:P, the local moments appear to be better described by a random-singlet phase.  This phase contains a diverging low-energy density of states that dominate thermodynamic measurements, and one might worry that their contribution to $\kappa$ would mask the metallic spinon contribution.  However these excitations are localized and inefficient at transporting energy.  A more careful estimate of the random-singlet thermal conductivity finds $\kappa_{\text{RS}} \sim T^4$ which is sub-dominant at low $T$.

\textit{Quantum Critical (QC) Scaling} --  Despite extensive experimental and theoretical effort, the quantum-critical (QC) behavior of electrical conductivity remains contentious and poorly understood.  The existence of a spin-liquid phase would have important implications for how QC scaling should be extracted and interpreted.  For $T=0$ and $n>n_c$, the system is a Fermi-liquid obeying the Wiedemann-Franz law: $\kappa/LT=\sigma$ (where the Lorenz number $L$ is a constant).  Since $\sigma$ vanishes at the transition while $\kappa/LT$ remains non-zero there must be a discontinuous jump in the $T=0$ electrical conductivity at the MIT.  In the slave-rotor description, this jump arises from the Ioffe and Larkin rule\cite{IoffeLarkin} that the electrical resistivity $\rho$ equals the sum $\rho = \rho_b+\rho_f$ of the resistivities of the bosonic rotors $\rho_b$ and spinons $\rho_f$ respectively.  Crossing the MIT at $T=0$, the bosons transition from a superfluid with $\rho_b=0$ to an insulator with $\rho_b=\infty$.  In contrast, the fermionic contribution, $\rho_f$ evolves smoothly through the transition, implying a non-universal jump in the zero-temperature conductivity.  Though superficially similar to Mott's early proposal\cite{Mott}, this jump in conductivity is unrelated to the idea of a ``minimum metallic conductivity".

Evidence against a discontinuous jump in conductivity in Si:P comes mainly from pressure tuning studies\cite{QCScaling} that show conductivity droping sharply but apparently continously to zero at the MIT.  However, determining whether one is truly accessing the asymptotic behavior near the QC point is very challenging, and the proper interpretation of the conductivity scaling remains controversial and poorly understood \cite{QCScalingControversy}.  For example, the pressure tuned experiments extract a conductivity scaling exponent $\nu=1/2$ that is incompatible with general exponent inequalities for a metal-insulator transition\cite{Chayes}, but could be explained as a thermally rounded version of the true $T=0$ conductivity jump.  In the spin-liquid scenario, as we will argue below, issues with extrapolating to $T\rightarrow 0$ and $n\rightarrow n_c$ are further exacerbated.   

\begin{figure}[ttt]
\begin{center}
\includegraphics[width=3.5in]{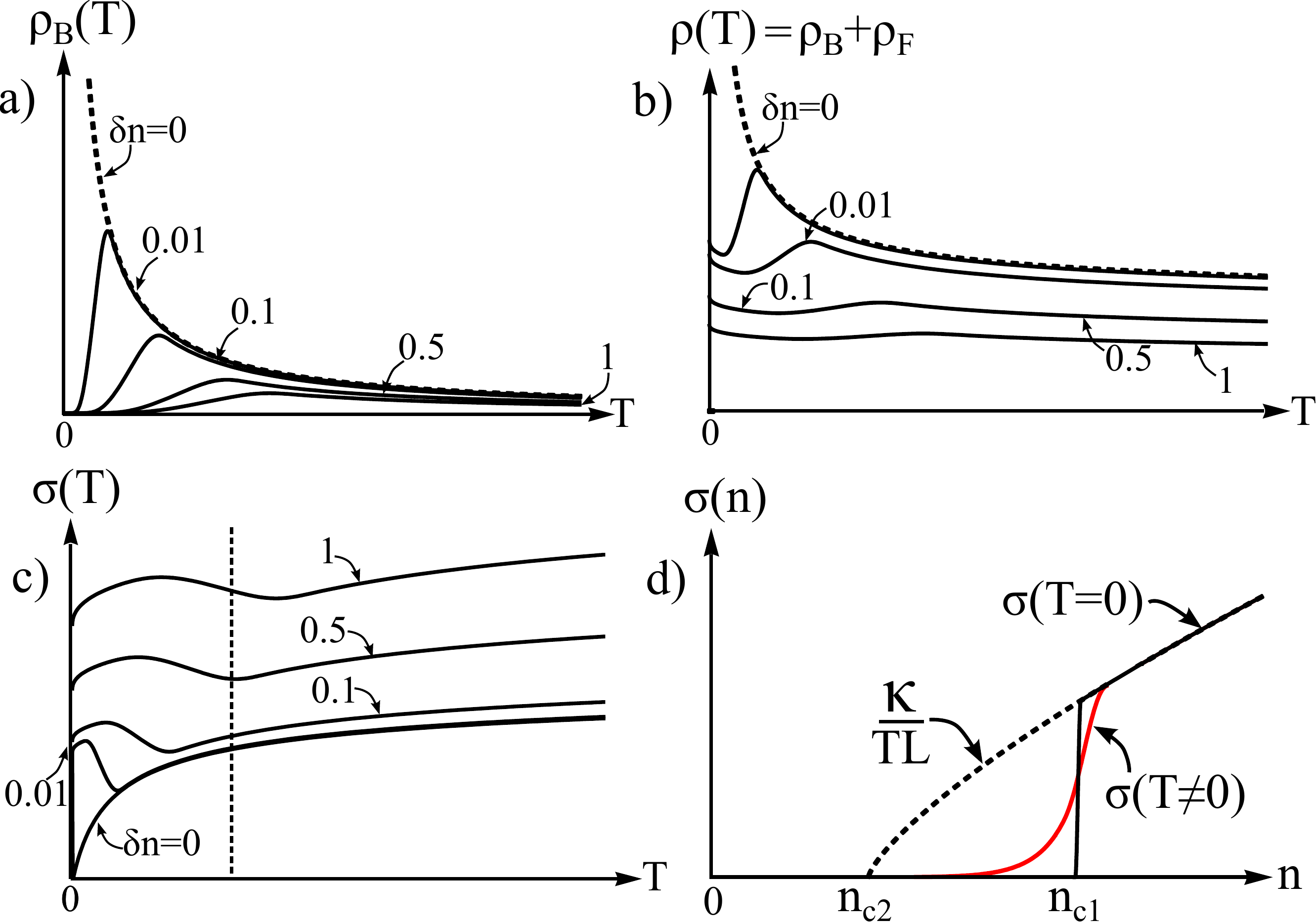}
\end{center}
\vspace{-.2in}
\caption{Quantum critical scaling (with $z=1$)\cite{zEndNote} of electrical conductivity $\sigma$ and linear--T coefficient of thermal conductivity $\kappa/TL$ ($L$ is the Lorenz number) near the MIT as a function of concentration and temperature.}
\vspace{-.2in}
\label{fig:ConductivityScaling}
\end{figure}

Near the quantum critical point (QCP) (i.e. $T\approx 0$, and $\delta n = n-n_c \ll n_c$), $\rho_b(T,\delta)$ obeys the quantum critical (QC) scaling for the disordered Bose-Hubbard model Mott transition as shown in Fig. \ref{fig:ConductivityScaling}a.  In the high-temperature critical regime where $T$ is the dominant perturbation away from criticality, $\rho_b(T)\sim T^{-1/z}$.  At lower temperatures $T<T^*\sim (\delta n)^{z/\nu}$, where $\delta n$ is the dominant perturbation from criticality $\rho_b(T)$ crosses over from the $T^{-1/z}$ to $0$ as superfluidity develops.  Here $z$ is the dynamical exponent for the disordered Bose-Mott transition with Coulomb interactions\cite{zEndNote}.

The spinon contribution to the resistance tends to a constant at zero temperature, due to the elastic scattering from disorder.  At finite temperature, there will also be non-constant contributions to $\rho_f$: $\rho_{\text{AA}}\sim \sqrt{T}$,\cite{AltshulerAronov} and $\rho_{\text{ph}}\sim T^{\sim 3-5}$, from interactions and phonons respectively.  The resulting electrical resistance, $\rho = \rho_b+\rho_f$, is depicted in Fig. \ref{fig:ConductivityScaling}b. for various $\delta n$ near the MIT.  The main feature here, is the resistance upturn due to the nearly--critical fluctuating bosons, which quickly disappears below $T<T^*$ as $\rho$ saturates to a non-universal constant set by $\rho_f$.  The corresponding $T$ dependence of  $\sigma = 1/\rho$ is shown in Fig. \ref{fig:ConductivityScaling}c.  Notice the discontinous jump in the very-low temperature conductivity between $\delta n\rightarrow 0^+$ and $\delta n = 0$.  As shown in Fig. \ref{fig:ConductivityScaling}d, this jump will be rounded at non-zero temperature, and could escape notice (consider, for example, if the lowest achievable temperature were indicated by the vertical dotted line in Fig. \ref{fig:ConductivityScaling}c).

The spin-liquid scenario outlined here suggests a very different scheme for extracting the QC behavior of conductivity, than that for a conventional localization transition.  Here, one should include only data for which the resistance saturates to a nearly constant value set by the spinon contribution.  In practice there is a minimum achievable value of temperature, $T_{\text{min}}$.  Consequently, this saturation region will disappear as the MIT is approached when $\delta n \lesssim T_{\text{min}}^{\nu/z}$.  Beyond this point, extrapolations based on the curvature of $\sigma$ would fail to capture the true $T\rightarrow 0$ behavior. 

The spin-liquid scenario will also complicate efforts to extract the critical scaling of $\sigma(n)$ near the MIT.  This difficulty is illustrated in Fig. \ref{fig:ConductivityScaling}d, which shows $\sigma(T=0,n)$.  As the concentration is decreased in the metal, the conductivity curves slowly towards an eventual localization transition at $n_{c2}$ (which may or may not occur).  However, in the present scenario, the Mott transition of the rotors intervenes at $n_{c1}>n_{c2}$ before the spinons localize.  In this case, extrapolations  of QC scaling based on a conventional Anderson transition from the metallic side would be misleading.

\textit{Discussion} -- In summary, we propose an alternative scenario to the Mott transition in doped semiconductors where the weakly insulating state is a spin-liquid with fermionic excitations. While such a transition has definite consequences for the quantum critical scaling of conductivity near the MIT, such quantum critical behavior is notoriously difficult to determine.  
%For example, previous critical scaling analyses of data have been plagued by controversies associated with finite temperature rounding as well as spatial correlations or finite size fluctuations in doping concentration. Furthermore, pressure tuning of $n$, the only existing method that allows for continous control of the MIT, leads to complicated and even non-mononotic dependence of $\sigma$ with pressure, making it difficult to establish critical scaling with confidence.  

Other possible signatures of spin-liquid behavior include sub-gap optical conductivity\cite{NgOpticalConductivity} in the insulator from gauge fluctuations and vanishing quasi-particle residue approaching the MIT (measurable by tunneling on the metallic side).  However, the former coexists with sub-gap conductivity from exciting weakly bound local moments, and the latter behavior will also be produced by a soft Coulomb gap (which will develop at the MIT)\cite{EfrosShklovskii}.  Consequently, such probes are indirect, and would require a detailed quantitative comparison.

Therefore, we suggest that the clearest test for spin-liquid behavior in doped semiconductors would come from a careful study of thermal transport near the MIT.  A spinon Fermi-liquid would lead to $\kappa \sim T$ for low $T$, which, if observed, would strongly indicate the presence of gapless fermionic excitations.

%Besides the thermal conductivity and critical scaling signatures discussed above, there will also be power-law optical conductivity for frequencies below the gap on the insulating side due to the emergent gauge field.  Also, as the MIT is approached from the metallic side in our scenario the quasi-particle residue $Z\rightarrow 0$, which could be observed in tunneling.  We caution, however, that both of these signatures are also present in more conventional scenarios due to a tail of localized states that contribute to optical conductivity in the Anderson insulator, and since the zero-energy tunneling density of states is suppressed to $0$ by long-range Coulomb interactions.  Therefore, to distinguish the spin-liquid scenario from the Anderson-Mott scenario using optical conductivity or tunneling would require a detailed quantitative comparison, which may be difficult.

\textit{Acknowledgements} --   MB acknowledges support by the Simons foundation, JM was supported in part by funds provided by the U.S. Department of Energy (D.O.E.) and in part by the Alfred P. Sloan Foundation under cooperative research agreement DE-FG0205ER41360, and TS was supported by NSF Grant DMR-1005434.

\appendix

\section{Appendix -- Disorder Averaged Properties of the Low--Energy Effective Slave--Rotor Theory}
In this appendix, we develop further, the properties of the low-energy effective field theory for the metal-insulator transition and spin-liquid phase within the slave-rotor theory.  In principle, a full treatment of disorder would need to account for correlations in the effects of local disorder fluctuations among the boson, fermion and the gauge field sectors.  We assume that the principal effect of such correlations is to produce local patches where the rotor and fermion are locally confined, i.e. to produce the dilute fraction of local-moments that are observed in each phase, but appear to decouple from the remainder of the system.  We expect that the formation of local-moments accounts for the effects of the rare-long tails of the disorder distribution, and that the remaining connected component of the system is reasonably characterized by treating disorder separately in each sector.   

In the boson-sector, the superfluid-insulator transition will occur in the presence of a random potential, and the resulting insulating phase will be a mixture of glassy puddles of superfluid which do not percolate, coexisting with a Mott-localized bulk.  In the spin-liquid scenario, the thermally conducting fluid of spinons (not including local moments) form a diffusive metal.  As shown below, the phase transition in the boson sector is not affected by the presence of gapless fermions or gauge fluctuations, and is identical to that of the ordinary dirty Bose-Hubbard model with random chemical potential and long-range Coulomb interactions\cite{Fisher}.  

\section{A. Irrelevance of Spinon and Gauge Fluctuations on the Slave--Particle Boson--Mott Transition}
\emph{Coupling to Fermions --} Including gauge fluctuations generically gives rise to a spinon-rotor density-density coupling of the form $\lambda_0 \delta n_f n_b$, where $\delta n_f$ is the deviation of $n_f$ from its average value.  Here we give a simple scaling argument that such a coupling does not alter the critical behavior of the boson--sector near the Mott transition.  Consider integrating out the spinons.  The leading order term in the effective action for the bosons will be of the form: $\lambda \sum_{\omega,\v{q}}\<\delta n_f(\omega,\v{q})\delta n_f(-\omega,-\v{q})\>|n_b(\omega,\v{q})|^2$.  In the transition to the spin-liquid, the spinon density-density correlator is diffusive and evolves smoothly across the transition: $\<\delta n_f(\omega,\v{q})\delta n_f(-\omega,-\v{q})\>\sim \frac{Dq^2}{|\omega|+Dq^2}$. 

After a momentum-shell renormalization--group (RG) step, integrating out modes with $q\in [\Lambda/s,\Lambda]$, with $s\gtrsim 1$, one rescales $q\rightarrow sq$ and $\omega\rightarrow s^z\omega$ to compare the new effective action to the original. Due to the random chemical potential provided by the spinon sector, the Boson Mott transition is in the same universality class whether one tunes through the transition either by changing chemical potential or hopping strength (this would not be true without the presence of a random chemical potential, where the Bose--Mott transition takes place at fixed density per site, with the random potential however, the density per site is only fixed on average).  For the chemical potential driven transition, under an RG step, the scaling part of the boson density rescales as $n_b(r,t)\rightarrow s^{d+z-1/\nu}n_b(r,t)$ where $\nu$ is the correlation length exponent. Equivalently, the fourier component rescales as $n_b(q,\omega)\rightarrow s^{-1/\nu}n_b(q,\omega)$.  

For $z<2$, the denominator of the diffusive fermion correlator is dominated by the $|\omega|$ term, and the density--coupling term scales as: 
$\lambda g\int d^dqd\omega q^{d-z}|n_b(q,\omega)|^2\rightarrow \lambda' s^{2(d-1/\nu)}\int d^dqd\omega q^{d-z}|n_b(q,\omega)|^2$, indicating that the coupling constant $\lambda$ rescales as $\lambda\rightarrow s^{2(1/\nu-d)}\lambda$.  Since $\nu d\geq 2$, $\lambda$ is irrelevant. Similarly if $z\geq 2$, the diffusive fermion correlator scales like a constant under RG, and the coupling constant $\lambda\rightarrow s^{2\nu-d-z}\lambda$ is again irrelevant.

\emph{Coupling to Gauge Fluctuations -- } The rotor--gauge field coupling generically takes the form $\int d^dr d\tau a_\mu j_b^\mu$, where $j_b$ is the boson current.  Integrating out the gauge-field at the RPA level generates a term of the form $g \int \dslash^d q\dslash\omega  G_a|j_b(\omega,\v{q})|^2$, where $G_a = \<|\v{a}(\omega,q)|^2\>$.  Current--continuity requires that $j_b(\omega,q)\sim \frac{\omega}{q}n_b(\omega,q)$, where $n_b$ is the \emph{total} boson density which scales like $n_b\sim 1/L^d$, indicating that $j_b(\omega,q)\rightarrow s^{-1}j_b(\omega,q)$ in each RG step.  As shown below, the gauge-field propagator scales like $G_a\sim (\omega+q^2)^{-1}$.  

For $z\leq 2$, under RG the gauge-fluctuation term rescales as $g \int \dslash^d q\dslash\omega  G_a|j_b(\omega,\v{q})|^2 \rightarrow g' \int s^{d+z}\dslash^d q\dslash\omega  s^{-z}G_a s^{-2}|j_b(\omega,\v{q})|^2$. Consequently $g\rightarrow gs^{2-d}$ flows to zero under RG and is irrelevant.  Similarly, for $z\geq 2$, the term rescales as $g \int \dslash^d q\dslash\omega  G_a|j_b(\omega,\v{q})|^2 \rightarrow g' \int s^{d+z}\dslash^d q\dslash\omega  s^{-2}G_a s^{-2}|j_b(\omega,\v{q})|^2$, indicating $g\rightarrow s^{4-(d+z)}g$, and the term is again irrelevant.

\section{B. RPA Effective Action for the Emergent Gauge Field in the Diffusive Spinon Metal}
Since the spinons and bosons have opposite charge under the emergent gauge field $a$, the gauge field couples to the currents as $S_{a-j} = \int d\tau d^3r \(j_\mu^b-j_\mu^f\)a^\mu$ (here and throughout, we work in imaginary time).  Therefore integrating out the spinon and boson fields within the RPA approximation, gives the following disorder-averaged effective action for the gauge field: 
\begin{equation} S_{\text{eff}}^{\text{(RPA)}} = \sum_{\omega,q} a_\mu(\omega,q)\[K_b^{\mu\nu}(\omega,q)+K_f^{\mu\nu}(\omega,q)\]a_\nu(\omega,q) \end{equation}
where $K_{f/b}^{\mu\nu}$ are the disorder--averaged current--current correlators (equivalently linear-response kernels) for the fermions and bosons respectively.

The temporal fluctuations of the gauge field are screened by the compressible fermions and become massive.  At the critical point and in the insulating phase the Boson conductivity vanishes, and consequently the gauge field dynamics are determined by the fermion response.  The disorder-averaged density-density part of the fermion electric response kernel is diffusive:
\begin{equation} K^f_{00} = \frac{2N(0)Dq^2}{|\omega|+Dq^2} \end{equation} 
where, we have expanded the diffusion ``constant" $D(\omega,q)$ (which generally has some $\omega$ and $q$ dependence, but does not vanish for $q,\omega\rightarrow 0$ outside of the disorder--localized phase) near $\omega=0=q$, and dropped the irrelevant higher order terms.  The other components of electric field response-function are related to $K^f_{00}$ by gauge invariance and charge conservation: $K^f_{0i} = K^f_{i0} = \frac{-\omega q_i}{q^2}K^f_{00}$, and $K^f_{ij} = \delta_{ij} \frac{-\omega^2}{q^2}K^f_{00}$.  Furthermore, we can identify $2N(0)D$ as the static uniform spinon--conductivity $\sigma_f$.  In addition, there is the usual diamagnetic response to fluctuating magnetic fields for $\omega\ll q$: $\chi_dq^2$, where $\chi_d$ is the Landau diamagnetic susceptibility of the spinons (whose average value is not altered by disorder).

Combining these considerations, and working in the Coulomb gauge ($\nabla\cdot\v{a}=0$ so that only the transverse gauge field $\v{a}_\perp$ remains) gives the following RPA action for the gauge-field $\v{a}$:
\begin{align} \label{eq:RPAGaugeAction} S_a^\text{(RPA)} = \sum_{\omega,q,\mu}\[\chi_d q^2+\frac{\sigma_f\omega^2}{|\omega|+Dq^2}\]|\v{a}_\perp(\omega,q)|^2
\end{align}
where $D = \frac{v_F^2\tau}{d}$ is the diffusion constant ($\tau$ is the disorder scattering time), and $\sigma_f$ is the spinon-conductivity. 

\section{C. Inelastic Scattering of Spinons from Gauge Fluctuations}
\begin{widetext}
Using the RPA expression for the gauge field propagator, one can find the leading (one-loop) self-energy for low-energy spinons near the Fermi-energy:
\begin{align} \Sigma(i\omega,k) &= \int \dslash \omega \dslash^2q_\perp \dslash q_\parallel D_a(i\Omega,q)\(v_F \frac{q_\perp}{q}\)^2G_f\(i(\omega-\Omega),k+q\)
\nonumber \\
G_f(i\omega,\v{k}) &= \frac{1}{i\omega-\xi_{\v{k}}+\frac{i}{2\tau}\sgn\omega}
\nonumber \\
D_a(\Omega,q) &= \frac{1}{\chi_D q^2+\frac{\sigma_f\Omega^2}{|\Omega|+Dq^2}}
\end{align}
\end{widetext}
In the fermion Green's function $\xi_{k+q}\sim v_Fq_\parallel +\mathcal{O}(q^2)$, indicating that $q_\parallel \sim \Omega$. Consequently one may approximate $q_\parallel^2\ll q_\perp^2$ in the gauge field propagator, making $D_a$ a function of $q_\perp$ only.  Furthermore, in this approximation, the current vertex $\(v_F \frac{q_\perp}{q}\)^2\approx v_F^2$.

Performing the $q_\parallel$ integral then gives $N(0)\sgn(\omega-\Omega)$ independent of the disorder scattering time $\tau$.  Considering the case of $\omega>0$ for definiteness, this limits the range of $\Omega$ integration from $0$ to $\omega$.  Since the dominant contributions come from $\Omega\sim q_\perp^2$, one finds that the spinon self-energy due to diffusively--screened gauge fluctuations scales like:
$\Sigma(i\omega) \sim \omega\log(1/\omega)$
Continuing to real-time one finds the inelastic gauge field scattering rate scales like $\text{Im}\[\Sigma^R(\omega)\]\sim \omega$, or equivalently $\tau_{\text{inelastic}}^{-1}\sim T$.  At low-temperature, this inelastic scattering is clearly sub-dominant compared to the elastic impurity scattering.


\begin{thebibliography}{99}
\bibitem{RSKappa}
The thermal-conductivity of a random-singlet phase of local moments is found to follow $\kappa_{\text{LM}}\sim T^{p'}$ with $p'\approx 4$ in 3D, and is also subdominant at low T (ACP -- Unpublished).
\bibitem{zEndNote}
In 3D $z$ need not be 1, but $z\approx 1$ is appropriate when the energetics are dominated by the long-range Coulomb interaction, which is plausible in the vicinity of the charge insulator.  Here we illustrate $z=1$ for definiteness, but stress that a different value will not qualitatively affect our results.
\end{thebibliography}
\end{document}